# Droplet microfluidics to prepare magnetic polymer vesicles and to confine the heat in magnetic hyperthermia

Damien Habault[1,2], Alexandre Déry[1,2], Jacques Leng[3,4], Sébastien Lecommandoux[1,2], Jean-François Le Meins[1,2], and Olivier Sandre[1,2]

[1] Université de Bordeaux, LCPO, UMR 5629, F-33600 Pessac, France
[2]CNRS, Laboratoire de Chimie des Polymères Organiques, UMR 5629, F-33600 Pessac, France
[3]Université de Bordeaux, LOF, UMR 5258, F-33600 Pessac, France
[4]CNRS, Rhodia, Laboratory of Future, UMR 5258, F-33600 Pessac, France

**In this work, we present two types of microfluidic chips involving magnetic nanoparticles dispersed in cyclohexane with oleic acid. In the first case, the hydrophobically coated nanoparticles are self-assembled with an amphiphilic diblock copolymer by a double-emulsion process in order to prepare giant magnetic vesicles (polymersomes) in one step and at a high throughput. It was shown in literature that such diblock copolymer W/O/W emulsion droplets can evolve into polymersomes made of a thin (nanometric) magnetic membrane through a dewetting transition of the oil phase from the aqueous internal cores usually leading to "acorn-like" structures (polymer excess) sticking to the membranes. To address this issue and greatly speed up the process, the solvent removal by evaporation was replaced by a "shearing-off" of the vesicles in a simple PDMS chip designed to exert a balance between a magnetic gradient and viscous shear. In the second example, a simple oil-in-oil emulsion chip is used to obtain regular trains of magnetic droplets that circulate inside an inductor coil producing a radio-frequency magnetic field. We evidence that the heat produced by magnetic hyperthermia can be converted into a temperature rise even at the scale of nL droplets. The results are compared to heat transfer models in two limiting cases: adiabatic *vs.* dissipative. The aim is to decipher the delicate puzzle about the minimum size required for a tumor "phantom" to be heated by radio-frequency hyperthermia in a general scope of anticancer therapy.**

## I. Introduction

*MAGNETIC VESICLES – especially those involving polymers either as protective shell or as membrane – constitute a major class of magnetic carriers with potential applications in translational research and nanomedicine. This paper proposes a generic way to marry them with the ever growing domain of microfluidics, both for the production of magnetic polymer vesicles and for the evaluation of their capability for heat production by radiofrequency magnetic field hyperthermia.*

### *A. Magnetic vesicles*

Around 40 years ago W. Helfrich, one of the pioneers of liposomal studies, already induced an ellipsoidal deformation of pure lipid vesicles playing on their diamagnetic anisotropy under an applied magnetic field strength $\mu_0 H$ of several T [1]. It was only more than twenty years later that C. Ménager *et al.* proposed a simple lipid rehydration method to encapsulate a superparamagnetic suspension (aqueous ferrofluid) inside the lumen of giant synthetic liposomes and obtain deformations under field strengths of 50 mT at most [2]. It has to be noted that single nanoparticles can also be wrapped independently by a phospholipid bilayer either naturally in the so-called "magnetosomes" originating from magnetotactic bacteria [3] or artificially with synthetic lipids and iron oxide nanoparticles [4]. Even though these systems sometimes referenced as "magnetoliposomes" can exhibit outstanding properties for magnetic hyperthermia or magnetic guiding, they obviously cannot deform as a function of the magnetic field intensity, because their core is totally filled by a solid iron oxide crystal. On the contrary, true magnetic liposomes are easily deformable because their membrane that encloses a given water volume possesses a large surface excess attested by large thermal fluctuations. Initially flaccid, such a vesicle filled by an ionic ferrofluid evolves either towards a prolate or an oblate ellipsoidal shape depending on the ionic strength, a behavior fully understood by taking into account the balance of magnetostatic, bending and electrostatic energies (within the approximation of Debye-Hückel theory) [5]. Validated by direct observation of giant magnetic liposomes under magnetic field, the same model was applied for the ellipsoidal deformation of "magnetic endosomes", the vesicles produced naturally by biological cells internalizing charged magnetic nanoparticles, as evidenced by cryo-TEM [6].

Analogous to liposomes by their structure, polymersomes are made of amphiphilic block copolymers self-assembled into closed membranes with hydrophobic thicknesses two-to-ten times larger than phospholipid bilayers [7]. Like liposomes, they can encapsulate hydrophilic magnetic nanoparticles inside their internal aqueous compartment [8]. But due to its large thickness, their membrane can also be loaded by magnetic nanoparticles once these are grafted with a suitable hydrophobic coating [9]. Small angle neutron scattering (SANS) experiments enabled evidencing anisotropic scattering patterns on sub-micron sized polymersomes under a magnetic field strength of 100 mT for polymersomes with hydrophobic blocks made of soft poly(butadiene) [10] or of more rigid and semi-crystalline poly(trimethylenecarbonate) blocks [11], the hydrophilic blocks being poly(L-glutamic acid) in both cases. However, the typical scale probed by SANS remaining below 100 nm, it was not possible to extract the exact shape of the

Corresponding author: O. Sandre (olivier.sandre@ipb.fr).
This paper has supplementary downloadable multimedia material available at http://ieeexplore.ieee.org or provided by the authors at www.dropbox.com/s/v32ji012vv9vrli/TMAG-sandre-2221688-mm.zip?dl=0. This includes ten experiment videos, which show the double-emulsion droplets and the magnetic polymer vesicles. This material is 33.3 MB in size.

polymersomes under field, but only to detect a variation of the membrane thickness described as a decrease by ~33% near the magnetic poles and an increase by ~50% near the equator. In theory, the expected deformation of a soft magneto-elastic shell under field is exactly reverse (elongation of the outer size of the vesicle in the field direction and contraction across the field) [12]-[13], although the modeling of "magnetostriction" remains a challenge for elastic materials [14]. Therefore a reproducible protocol to prepare giant magnetic polymersomes – observable in real time by optical microscopy – is highly suitable for deciding whether polymersomes with magnetic nanoparticles embedded in their membranes under an applied magnetic field deform into ellipsoids and in that case whether the eccentricity is positive (prolate) or negative (oblate). This would enable extrapolating the behavior of sub-micron sizes' polymersomes currently under active study as multi-functional drug carriers with externally triggered release and negative contrast enhancement for MRI [11],[15], by envisaging also a magnetic actuation under a static magnetic field.

### B. *Microfluidics to prepare vesicles*

Among the other methods to generate giant vesicles (film rehydration or reverse-phase evaporation (REV) [2], multiple emulsion [16], electro-formation [17], emulsion-centrifugation also called "droplet interface crossing encapsulation" (DICE) [18]…), the double-emulsion technique presents the advantage of controlling precisely the composition of the two aqueous phases (internal and external) and of the membrane. It consists in preparing a double water-in-oil-in-water W/O/W emulsion, where the oil is a solution of the lipid or the block copolymer in a good solvent (possibly a mixture) not miscible with water. When the solvent evaporates, the amphiphiles self-assemble into closed membranes encapsulating the internal water phase. However, this scenario is rather ideal and several problems can arise before the vesicles eventually form. At first the formulation of a stable W/O/W emulsion is delicate and can necessitate co-surfactants and osmotic agents to prevent the coalescence of dispersed aqueous droplets into the continuous phase. Then the main drawback is the broadness of sizes of the emulsion droplets, and thus of the vesicles in the end. Therefore microfluidics was proposed both to control the sizes and to study the mechanism of vesicles formation via spatial control and temporal monitoring [19]-[20]. A microfluidic version of the DICE method was also proposed [21], offering to combine a high encapsulation yield to a high throughput.

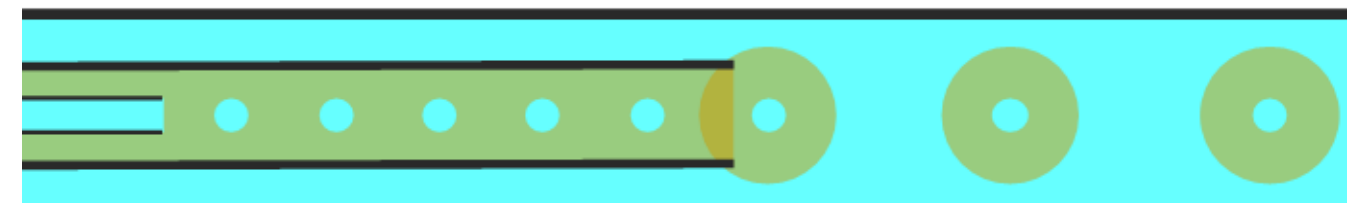

Scheme 1. Coaxial flows geometry of a microfluidic chip to produce a double W/O/W emulsion for the production of giant vesicles.

Different natures of copolymers and solvents have been reported to produce giant polymersoomes by double-emulsion, *e.g.* toluene/THF mixture for hydrophobic blocks made of poly(butylacrylate) [22], toluene/chloroform for poly(styrene) or poly(butadiene) [19] and poly(lactic acid) blocks [20],[23], and hexane/chloroform for poly(lactic acid) blocks [24],[25], the hydrophilic ones being often poly(ethylene oxide). But in all cases, there is a narrow concentration range of polymer (*e.g.* 0.5 – 1 wt%) that is required for the stability of the emulsion. However, this amount is one to two orders of magnitudes higher than the polymer concentration needed theoretically to cover the internal droplets by a molecularly thin membrane (either a bilayer for diblock or a monolayer for triblock copolymers). Therefore solid "acorn-like" structures containing this copolymer excess usually remain stuck to the final polymersomes produced by the microfluidic double-emulsion technique [19]. More precisely, Shum *et al.* showed that the transformation from double-emulsion droplets into vesicles is not continuous with a slow decrease of the hydrophobic shell thickness, but rather involves an abrupt transition: along evaporation of the more volatile component in the solvent (also slightly soluble in water in the case of chloroform), there is a critical fraction of this solvent below which the copolymer dewets from the internal aqueous droplet [24], which gives rise to the acorn-like structure. The authors also suggest to generate this dewetting intentionally as soon as possible in the chip: thus by applying a hydrodynamic shear on the dewetted emulsion drop (an oily droplet containing the copolymer excess still adhering to the internal droplet, this being wrapped by a thin membrane as seen on Fig. 4 in [24]), the vesicles can be produced by "shearing-off" much more rapidly than by waiting for the full evaporation of the solvent. Using the same flow-focusing chip, C. Ménager *et al.* obtained giant magnetic liposomes and polymersomes with the magnetic nanoparticles located either in the aqueous cores or in the membranes [26], but did not discuss about the deformations under a static magnetic field. In the present work, we show a chip applying a magnetic force on the magnetic oil, leading to a particularly efficient control of the shearing-off and the collection at the outlet of the chip of magnetic polymersomes with almost no copolymer excess.

### C. *Magnetic hyperthermia: between dreams and physical limitations*

For almost two decades [27], magnetic fluid hyperthermia (MFH) has strongly stimulated the imagination of physicists, biologists and materials scientists, who thought about applying MFH in anticancer therapies. This dream has become a reality, since MFH has passed a phase II clinical trial on humans, in combination with classical radiotherapy [28] and is now authorized on market. In parallel of this direct use of MFH to destroy malignant cells (thermal ablation), a more gentle use of MFH consists in transferring the heat produced to a thermo-sensitive drug carriers. This approach was developed initially with phospholipid vesicles, which bilayer was decorated with hydrophobically coated magnetic nanoparticles, either with freely rotating magnetic moments (maghemite) [29],[30],[31] or blocked ones (cobalt ferrite) [32]. Drug or model dyes were released under an oscillating magnetic field at radiofrequency (100 kHz – 1 MHz) in the former case. In the latter case (blocked moments), a lower frequency should be used (~kHz) involving solid rotation of the nanoparticles in the vicinity of the membrane, a reminiscence of the "molecular drill" effect proposed long ago by P-G. de Gennes *et al.* [33].

In the field of MFH-triggered release, we can distinguish between studies reporting a macroscopic temperature rise (which has an effect on a model drug diffusion rate or on a thermal transition of a polymer) and those that evidence an enhanced release rate without a detected temperature change. An example of the first class is the release into cells of a cytotoxic drug, doxorubicin (DOX), from magnetic alginate micro-beads sitting just underneath the cell culture [34]. Another one involves worm-like micelles of the commercial copolymer Pluronic F-127 releasing lysozyme through a sol-gel transition induced by MFH [35]. However, several studies recently evidenced that a macroscopic temperature rise was not a prerequisite for MHF-induced drug release: the enhanced release of DOX from sub-micron magnetic polymersomes [11] or the cytotoxicity of polysaccharide-coated iron oxide nano-particles functionalized with an appropriate antibody targeting cancer cells [36]. These observations give a new hope for the application of MFH to drug delivery systems, despite the physical limitations that any scientist involved in this domain should keep in mind: on the one hand the minimum amount of magnetic material, deduced from the specific heating power of the magnetic nanoparticles (SHP in W/g, normalized by the weight of magnetic material) arising from the magnetic relaxation mechanisms themselves [37], [38], [39]; on the other hand the minimum size of the heated sample, due to heat losses by thermal conductivity into the ambient medium [40], which necessitates to design very properly the geometry and size of the experiments and to work in adiabatic conditions for the SHP measurements [41]. Until now the thermal ablation is possible only on primary brain tumors (typical size ~cm) fed with an iron oxide concentration above 30 mg·mL$^{-1}$ [28]. In future, secondary tumors and metastases could be reached by targeting magnetic carriers. However this goal can be achieved only by a careful design of the systems and the knowledge of the required distribution of magnetic materials (quantity and spatial distribution) that can produce the minimal temperature increment in a given volume necessary to induce an effect on the carrier (*e.g.* reaching the melting temperature of a polymer membrane). Therefore we designed a microfluidic setup to measure temperature locally, *i.e.* at 33 nL scale, on a train of ~400 µm diameter droplets circulating inside an inductor coil.

## II. Materials and Methods

### A. Magnetic fluid

In order to prepare a magnetic fluid with good susceptibility and proper dispersion state, we used the iron salts alkaline co-precipitation route in water also called Massart's process [42]. By measuring the magnetization curve by vibrating sample magnetometry (VSM) and a Langevin fit convolved with the Log-normal law [43], we got the distribution of diameters of the magnetic nanoparticles (MNPs): $d_0$= 8.1 nm and $\sigma$=0.26. The iron concentration (1.6 mol·L$^{-1}$) being measured by UV-Vis spectroscopy using a calibration curve, the plateau of the magnetization curve leads to a specific magnetization for these nanoparticles $m_S$=2.9×10$^5$ A·m$^{-1}$. Initially in water with dilute nitric acid (pH=1.4), the suspension was transferred into an organic solvent by coating with an appropriate surfactant: either classical oleic acid for dispersion in cyclo-hexane or the commercial phosphoric ester Beycostat™ NB09 for dispersion in dichloromethane. In both cases, the grafting reaction was performed under gentle heating (60°C) at 20 mol% ratio of surfactant relatively to iron in alkaline condition, followed by several washing steps in methanol. The suspension coated with Beycostat NB09 has a hydrodynamic diameter $d_H$=30 nm (PDI=0.18) in $CH_2Cl_2$. The MNPs stabilized by oleic acid in cyclo-hexane have a $d_H$=28 nm (PDI=0.13), as measured on a Vasco DL135 (Cordouan Techno., Pessac, France), a DLS setup at a scattering angle of 135°C dedicated to strongly absorbing fluids like magnetic fluids, with a diode laser at a 650 nm wavelength working at 50% of full power (65 mW). When used for microfluidics, the stock suspensions were adjusted to 0.1 wt% in the corresponding solvent or mixture.

### B. Amphiphilic block copolymers

Two batches of poly(butadiene)-*b*-poly(ethylene oxide) were purchased from Polymer Source Inc. (Dorval, Canada): $PBD_{6.5k}$-*b*-$PEO_{3.9k}$ (#P4753-BdEO) and $PBD_{2.5k}$-*b*-$PEO_{1.3k}$ (sample #P9095-BdEO). According to the provider, the molar-mass dispersities are 1.1 and 1.09, respectively. The PEO blocks are semi-crystalline ($T_m$=48°C). At room temperature, the PBD blocks are well above Tg (-31°C for $PBD_{2.5k}$ and -20°C for $PBD_{6.5k}$, respectively), which can be ascribed to their high amount of 1,2-diene addition (89% for $PBD_{2.5k}$ and <80% for $PBD_{6.5k}$, respectively) favoring a branched structure. However, we expect $PBD_{2.5k}$ and $PBD_{6.5k}$ to be respectively slightly below and above the critical entanglement mass, leading to distinct viscoelastic behaviors of the membranes.

### C. Microfluidic production of W/O/W emulsion droplets

The original device proposed by D. Weitz *et al.* [19] for double-emulsion production consists in two tapered glass tubes fitting on both sides of a third capillary with a square section: this trick enables a perfect alignment of the symmetry axes of the three capillaries, and thus perfect co-axial flows. An alternative which does not guarantee such a perfect alignment but is easier to implement was developed in parallel by several groups [44],[45],[46]. This simpler chip is based on the imbrications of three fused silica capillaries by adjusting them into commercial sleeves and T-junctions. For the double-emulsion setup, we used three capillaries (Polymicro Tech.) thereafter denoted according to their size SC for small (*ID*=40 µm, *OD*=110 µm), MC for medium (*ID*=150 µm, *OD*=360 µm) and LC for large (*ID*=540 µm, *OD*=670 µm), as depicted on Scheme 1.

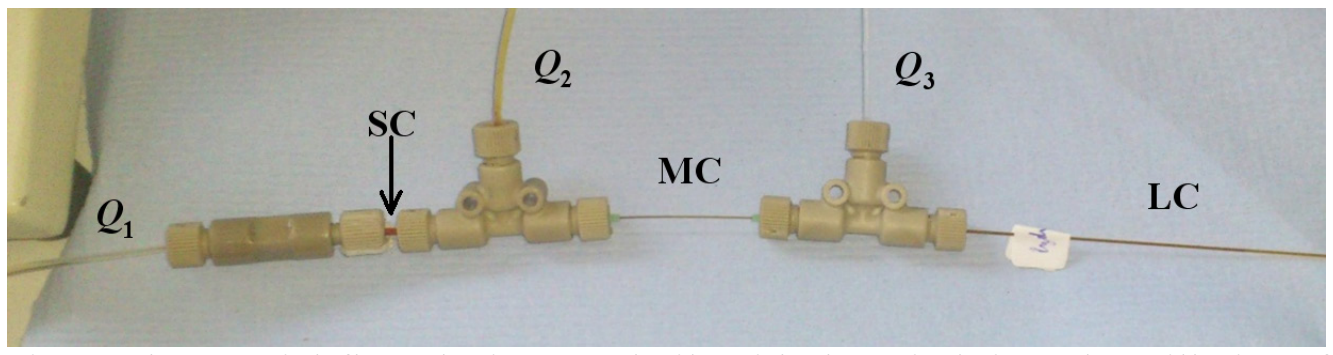

Fig 1. The coaxial flow device was built with three imbricated capillaries of increasing diameters (SC, MC, LC) by using commercial tubing sleeves and T-junctions (Upchurch Scientific), the smallest one (SC) being hidden inside its sleeve. The flow-rates (respectively $Q_1$, $Q_2$, and $Q_3$) were controlled by precision syringe pumps.

Each capillary was inserted in one corresponding sleeve: SC in F-237X, MC in F-242X, and LC in F-246X (Upchurch Sci). Then the whole setup is mounted as seen on Fig. 1 using the following references from Up Church Sci.: P-659 10-32 female-to-female luer connector, F-33 10-32 PEEK nut, P-235X 1/16 PEEK short nut, P-200X flangeless 1/16 nut, P-713 PEEK 1/16 Tee. The MC intended to be filled with the oily mixture was coated by a self-assembled monolayer (SAM) of octadecyltrichlorosilane (OTS) using a standard protocol [47], where carbon tetrachloride was replaced by chloroform. Each capillary was connected though a tubing to a syringe, which flow-rate was controlled independently by a pump (Harvard Apparatus). The outlet of the device (end of LC) was placed above collection vials (2 mL centrifuge tube, Ependorf™). The SC and LC were filled with aqueous solutions containing 2 wt% PVA (87-89% hydrolyzed, $M_w$=13000-23000 g·mol$^{-1}$) to prevent coalescence of the droplets and 50 ppm $NaN_3$ as bactericide. Besides viscosity enhancement, PVA also induces a large decrease of interfacial tension for all oil compositions, as seen on Table 1 and Fig. 2. In addition, the internal (SC) and external (LC) solutions contained 0.1 mol·L$^{-1}$ of a sugar (respectively sucrose and glucose), to ease the observation of the droplets or the vesicles by optical phase contrast, and to enable their sedimentation at the bottom of the collection vial.

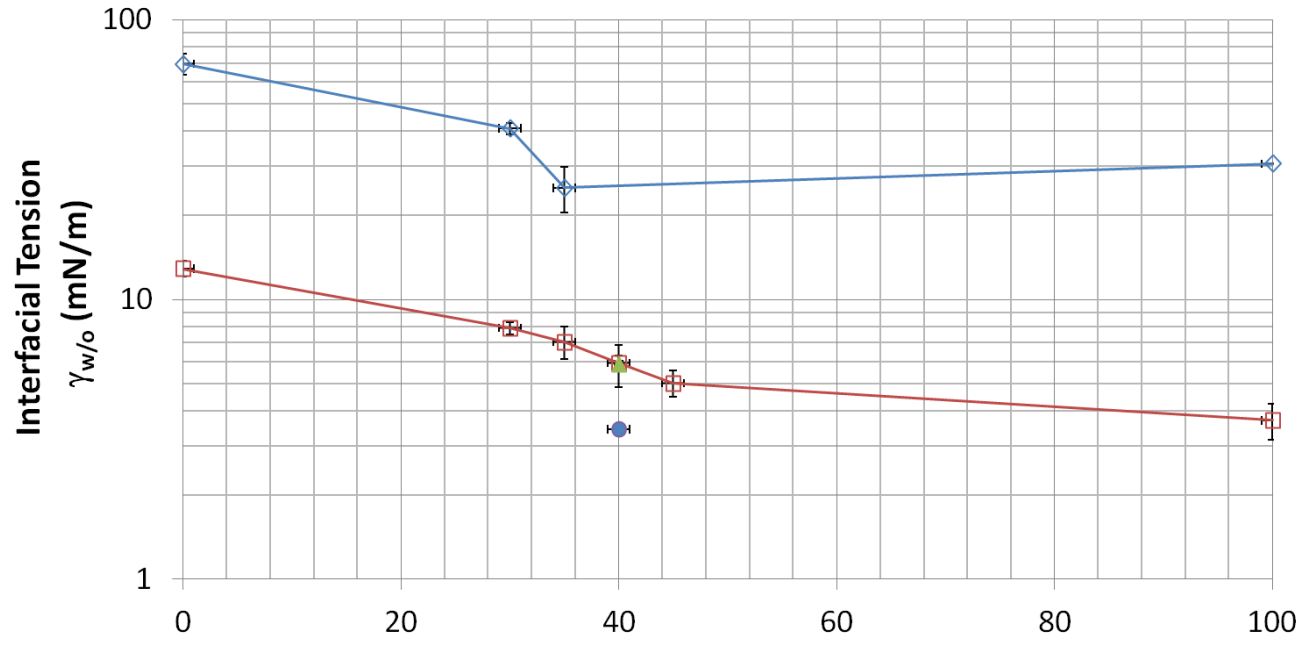

Fig 2. Variation of the interfacial tension between a chloroform/cyclo-hexane mixture of varying composition and with different aqueous phases: pure water (blue diamonds), 2 wt% PVA (red squares), 1 wt% $PBD_{6.5k}$-*b*-$PEO_{3.9k}$ (green triangle), 0.1 wt% MNPs coated by Beycostat (blue circle).

TABLE I
INTERFACIAL TENSIONS FOR VARIOUS COMPOSITIONS OF PHASES

| Aqueous solution | *cyclo*-hexane/chloroform (60:40) | Interfacial tension $\gamma_{w/o}$ (mN·m$^{-1}$) [a] |
|---|---|---|
| Pure water | Pure solvent mixture | 25.1±4.7 [b] |
| 2 wt% PVA | Pure solvent mixture | 5.9±0.4 |
| Pure water | 1 wt% $PBD_{2.5k}$-b-$PEO_{1.3k}$ | 5.1±0.2 |
| 2 wt% PVA | 1 wt% $PBD_{2.5k}$-b-$PEO_{1.3k}$ | 3.3±0.6 |
| Pure water | 1 wt% $PBD_{6.5k}$-b-$PEO_{3.9k}$ | 5.9±0.4 |
| 2 wt% PVA | 1 wt% $PBD_{6.5k}$-b-$PEO_{3.9k}$ | 2.3±0.3 |
| Pure water | 2 wt% Beycostat NB09 | 6.4±0.2 |
| 2 wt% PVA | 2 wt% Beycostat NB09 | 1.8±0.1 |
| Pure water | 0.1 wt% nanoparticles@Bey | 3.4±0.1 |
| 2 wt% PVA | 0.1 wt% nanoparticles@Bey | 3.3±0.3 |

This table illustrates the influence of the amphiphilic compounds on $\gamma_{w/o}$ for a given weight ratio of solvents in the oil phase (60:40).
[a] average of 5 measurements by the pendant drop method.
[b] this value was measured actually for *cyclo*-hexane/chloroform (65:35).

The intermediate capillary (MC) was filled by a mixture of chloroform and cyclo-hexane. In preliminary experiments, the proportions were varied from 30:70 to 50:50 to find the optimal weight ratio enabling production of W/O/W droplets, and in the end of vesicles. This ratio has a tremendous effect on the interfacial tension between water and oil, as plotted on Fig. 2. Then a weight ratio of 50:50 was chosen for all the microfluidic experiments. The idea consisted in choosing the chloroform content in the mixture just above the threshold of dewetting transition, so that the dewetting occurs rapidly inside the device (not later in the collection vial) [24]. In the case of $PLA_{5k}$-*b*-$PEO_{5k}$, this transition is located at 37 vol% chloroform (53:47 in weight). But for $PBD_{2.5k}$-*b*-$PEO_{1.3k}$, the transition must be at a slightly lower chloroform content, because for 34.5 vol% (50:50 in weight) the mixture is still a good solvent for the diblock copolymer, as attested by DLS: the hydrodynamic size of this 50:50 mixture with 1 wt% $PBD_{2.5k}$-*b*-$PEO_{1.3k}$ and 0.1 wt% MNPs coated by oleic acid remains at a low value, $d_H$=38 nm (PDI=0.08).

### D. *Chamber for magnetic-assisted recovery of vesicles*

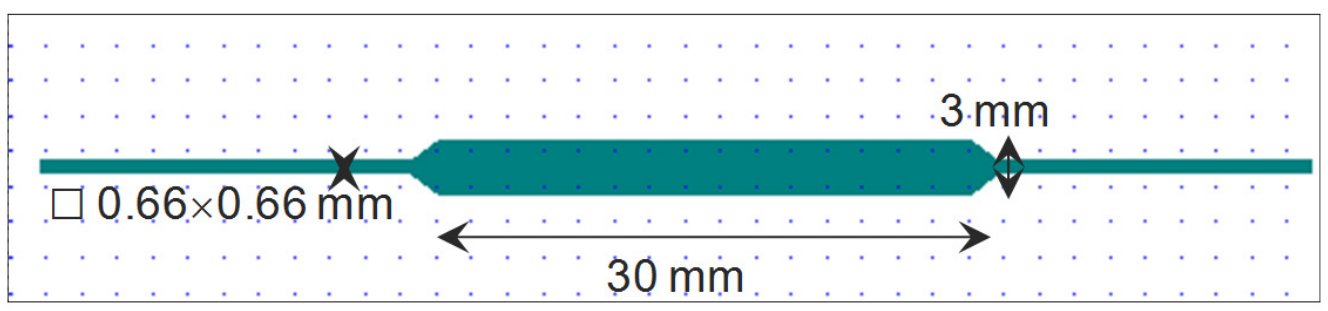

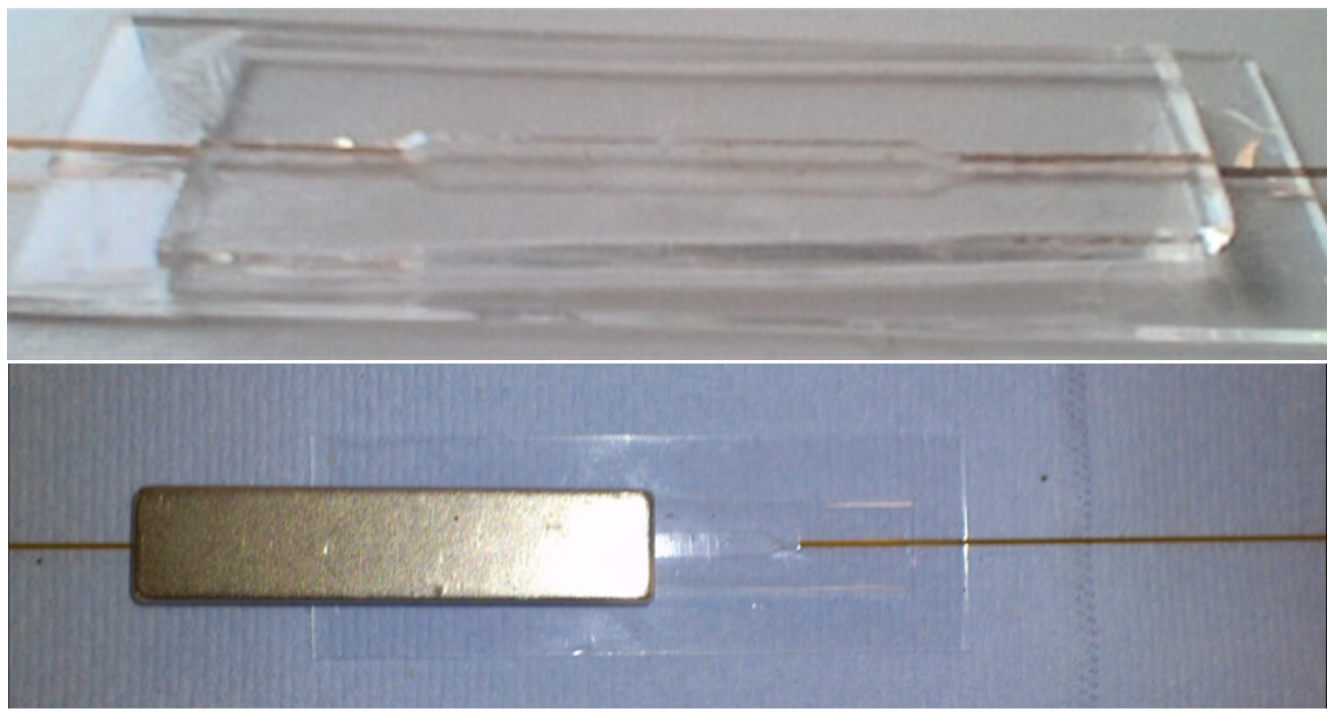

Fig 3. Dimensions and pictures of the PDMS chamber designed for magnetic-assisted shearing-off. The entry on the left is connected to the large capillary (LC) of the coaxial flows device producing the double-emulsion, the outlet on the right goes into a collection vial. The strong permanent magnet is simply placed on top of the channel on the left side, at a short distance of the channel.

The combination of microchannels and magnets is reported in the literature, in order to sort particles or biological cells according to their sizes and their magnetic content [48]. Here we make a very basic design. A strong NdFeB magnet (slab 56×13×6 mm, N35 grade, Calamit Inc.) is simply placed at a close distance (~0.5 mm) above the ceiling of a large channel pictured on Fig. 3. This chamber was fabricated using classical PDMS technology. The mould was made by microlithography using the design of Fig. 3 as mask. The difficulty arises from a rather large height of the pattern (660 µm), leading to adapt the development time of the photo-resist. Chambers of varying distance between the upper surface of the channel and the magnet were obtained by varying the amounts of PDMS (Sylgard 184) poured into the mould and cured at 80°C, after degassing. Before use, the chambers were made hydrophilic by flowing 5 mL of 1 wt% BSA solution in 30 min.

### E. Device for circulation inside an induction coil

Another device was built for a completely different purpose. The idea was to circulate a train of magnetic fluid droplets within an immiscible fluid of thermal conductivity much lower than water ($k_{water}$≈0.6 W·K$^{-1}$·m$^{-1}$). In order to detect a pure MFH effect with no contribution of electric conductivity (eddy currents), the magnetic fluid coated by Beycostat™ was chosen for the droplets instead of an aqueous magnetic fluid. The continuous phase was perfluorohexane $C_6F_{14}$ (Fluorinert™ FC-72), with the characteristic values given by the provider (ABCR GmbH, Karlsruhe, Germany): $k_{C6F14}$=0.057 W·K$^{-1}$·m$^{-1}$, $C_p$=1100 J·kg$^{-1}$·K$^{-1}$, $\eta$=6.4×10$^{-4}$ Pa·s, $\rho_{C6F14}$=1680 kg·m$^{-3}$, and $\theta_{boil}$=56°C. The emulsion was obtained with a device made of two capillaries of same sizes than MC and LC described before. MC was treated with OTS as previously. LC was coated with 1H,1H,2H,2H-perfluorodecyltrichlorosilane (FAS-17, provided by Alfa Aesar) improving the wetting of glass by fluorinated compounds [49].

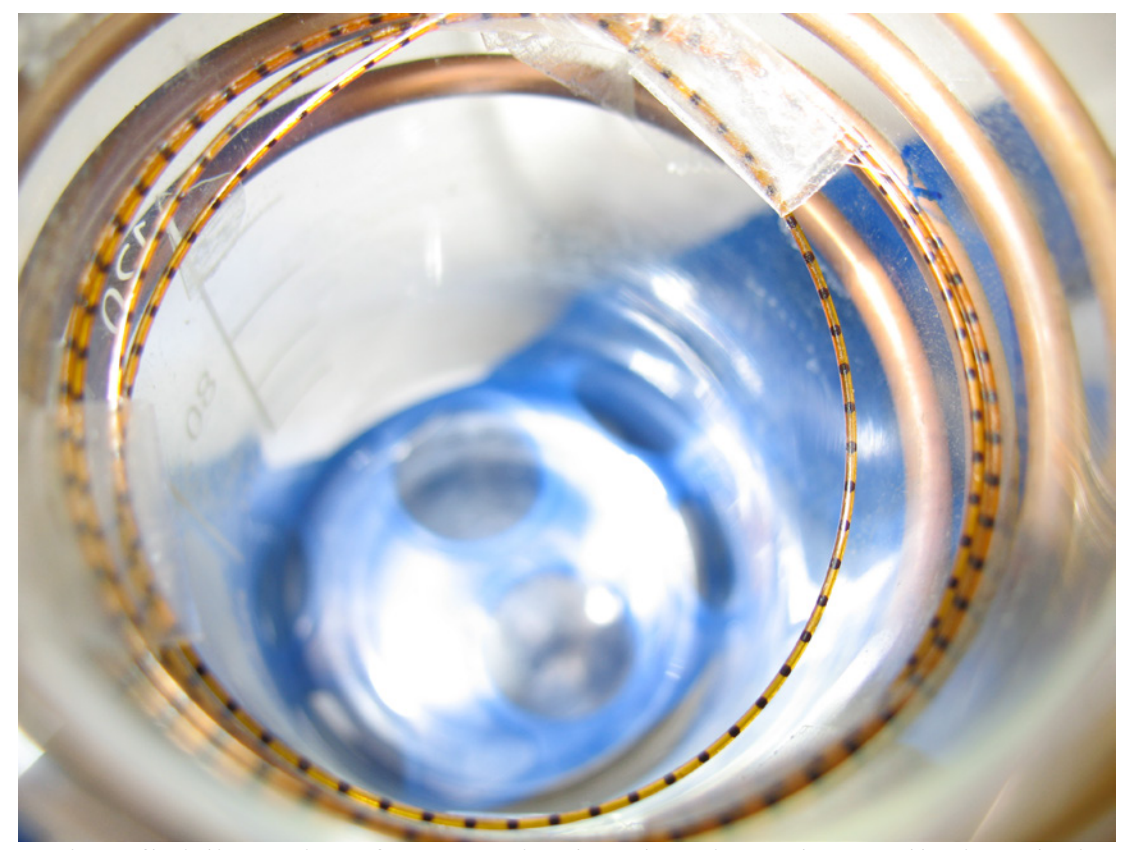

Fig. 4. Microfluidic train of magnetic droplets in a thermally insulating oil to confine the heat produced by RF hyperthermia. Here the flow-rates $Q_2/Q_3$ are respectively 50/150 µL·min$^{-1}$ thus the magnetic fluid occupies $Q_{MF}$=25% v/v.

### F. Magnetic radiofrequency hyperthermia setup

A commercial induction generator (Tireless 3 kW, 800 kHz nominal frequency) was purchased from Seit Elettronica, Italy. The coil with a diameter of 5 cm and a height of 3.2 cm contains 4 turn. The maximum current being 90 A according to the provider, the calculated field induction is $B$=14 mT. A measurement under load gave 755 kHz as actual frequency. A measurement of the slope of temperature *vs.* time on 1 mL of the magnetic fluid synthesized in this work lead to SHP=9±1 W/g. This value correlates perfectly with the reported value SHP=37 W/g for $\gamma$-$Fe_2O_3$ MNPs of same size distributions ($d_0$=8.0 nm and $\sigma$=0.21) at almost same frequency (700 kHz) but at twice stronger field (the SHP is expected to scale as $B^2$) [38]. A typical length $L$=1 m of the largest size capillary (LC) was bent and placed within the coil, as pictured on Fig. 4. The flow-rates of magnetic fluid and fluorinated oil were denoted respectively $Q_2$ and $Q_3$, by analogy with the previous device. The time of residence of the magnetic droplets depends on the length of LC and on the total flow rate through $\tau_{res}=\pi \cdot ID^2 \cdot L/4(Q_2+Q_3)$ (*e.g.* 69 sec for $L$=1 m and $Q_2+Q_3$=200 µL·min$^{-1}$ as on Fig. 4). Temperature could be followed directly inside the train of droplets using an optical fiber thermometer insensitive to magnetic fields (OTG-M360, Opsens, QC, Canada) inserted inside LC, as seen on Fig. 5.

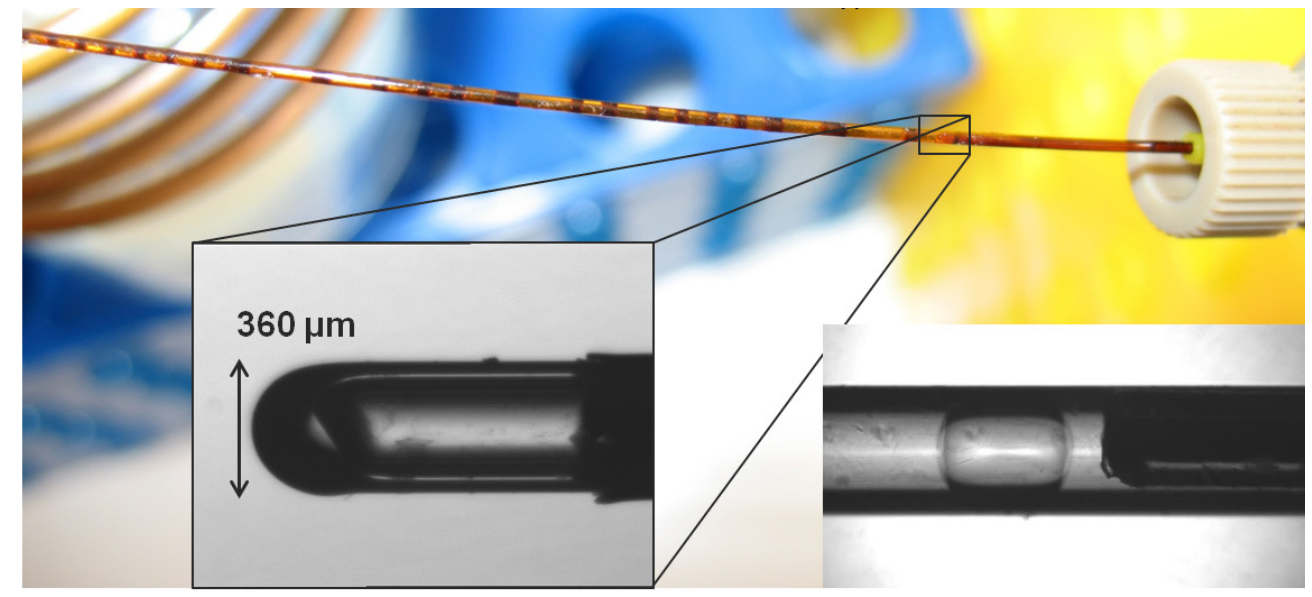

Fig. 5. Measurement of $T$ within the droplets by insertion of an optical fiber thermometer. The droplets can flow along the probe without trouble (inset).

## III. Results and Discussion

### A. Microfluidic production of giant magnetic polymersomes

#### 1) Production of a stable double-emulsion

The first step was to obtain magnetic double-emulsion droplets by adjusting both the compositions and the flow-rates. At first it was impossible to get a stable double-emulsion for a copolymer concentration lower than 0.5 wt%. Obviously the strong decrease of interfacial tension $\gamma_{w/o}$ due to the copolymer amphiphilicity (Table 1) is a prerequisite for the stability of the internal W/O droplets. The size of the droplets (W and O) and the number of encapsulated W droplets per O drop follow scaling laws as a function of flow-rates that have been studied in a previous work [46]. Here we just verified general trends: the flow-focusing effect varies with the ratio of the flow-rates, which determines the size of the droplets. Thus the diameter of the W droplets ($D_{Int}$) decreases when increasing $Q_2/Q_1$, and the diameter of the O droplets ($D_{Ext}$) increases when $Q_3/(Q_1+Q_2)$ decreases, *e.g.* from 6 to 3 as seen on Fig. 6. These ratios also determine the regime between continuous jets and droplets, together with the number of W droplets per O droplet.

#### 2) Recovery of vesicles by magnetic-assisted shearing-off

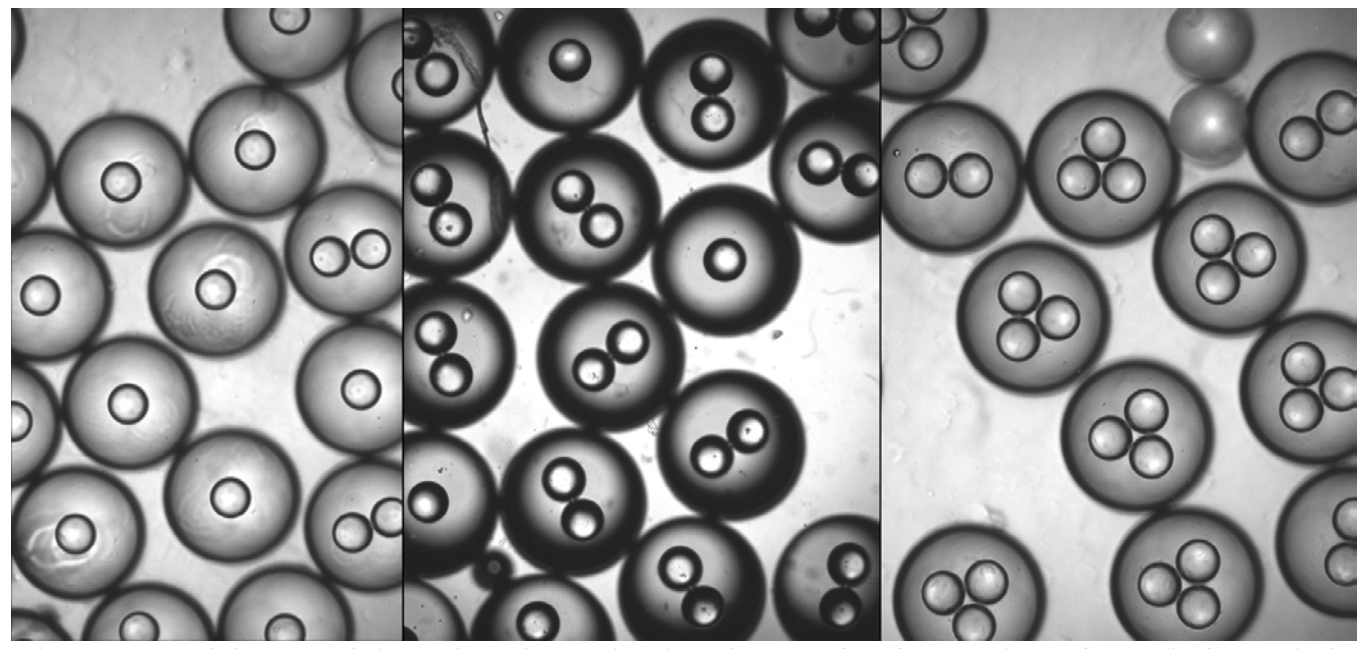

Fig. 6. Double-emulsion droplets obtained at $Q_1/Q_2/Q_3$ ratios, from left to right (µL·min$^{-1}$): 0.5/49.5/300, 0.5/49.5/200, and 0.5/49.5/150. $D_{Int}$=173±3 µm is constant ($Q_1/Q_2$=100 fixed) and $D_{Ext}$ varies from 527±6 µm (mainly single W droplets) to 595±6 µm (mainly doublets) and 607±4 µm (mainly triplets).

In preliminary experiments, double-emulsion droplets were collected in tubes and let at rest overnight in order to reach the complete evaporation of the solvents. Magnetic polymersomes did form sometimes, but this protocol was erratic: the vesicles were not perfect spherical shells; they were always sticking to a polymer precipitate (acorn-like in some cases). Thus we develop the magnetic-assisted shearing-off cell, described in Materials and Methods. A typical sequence is shown on Fig. 7.

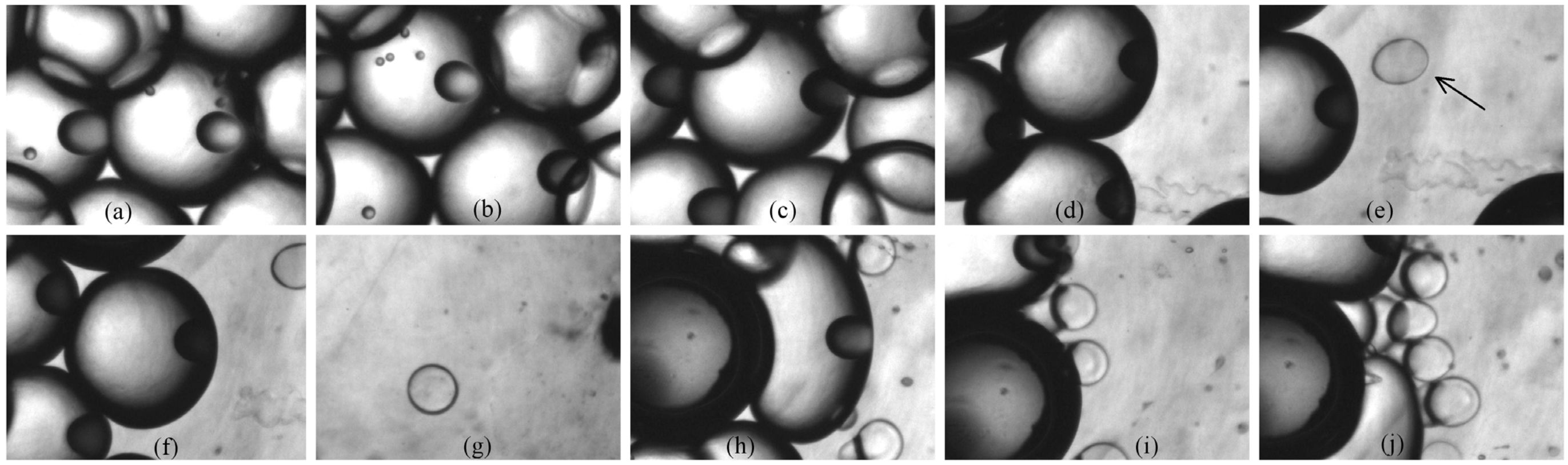

Fig. 7. “Magnetic-assisted shearing-off” of the polymersomes (time increases from (a) to (j).) The large O droplets loaded in magnetic fluid are attracted by the field gradient produced by the magnet (located above the channel, on the left side), while the internal W droplets are repelled (“magnetic holes” effect, (a) to (c).) Thus they accumulate on the right side of the aggregate (d). By flushing the chamber continuously, the stream induces a shear on the small W droplets, which eventually escape (e, arrow) and migrate to the outlet (f, g). As in the DICE [18] and c-DICE [21] techniques, the droplets move across the O/W interface (h-j).

A stable double-emulsion regime was obtained typically for $Q_1$=0.5-1 µL·min$^{-1}$, $Q_2$=30-50 µL·min$^{-1}$, $Q_3$=200-300 µL·min$^{-1}$. Immediately after placing the magnet on top of the chamber, the flows in SC and MC were switched off ($Q_1$=$Q_2$= 0) and $Q_3$ was decreased down to *e.g.* 60 µL·min$^{-1}$. When the magnetic O droplets had started to agglomerate below the magnet as observed by optical microscopy (Fig. 7), the aqueous flow-rate $Q_3$ was raised again by increments of 10 µL·min$^{-1}$ every 5-10 min, until internal W droplets (“magnetic holes”) repelled by the magnetic field gradient were accumulating on the edge of the magnetic aggregate. The external flow-rate $Q_3$ was again increased by increments of 10 µL·min$^{-1}$, until the W cores were pulled-out of the oil phase by the hydrodynamic shear stress. In all cases, $Q_3$ should not exceed 120 µL·min$^{-1}$, otherwise the shear becomes too large compared to the magnetic attraction, and vesicles become mixed with small O droplets in the collection vial.

*3) Observation of the giant magnetic polymersomes*

Immediately after magnetic-assisted shearing-off, the membranes of the vesicles were still swollen by a small amount of solvent; which apparently evaporated within 1 hour. The magnetic polymersomes exhibited a great regularity and monodispersity of diameters (Fig. 8). They were thus ready for observation and manipulation under a magnetic field applied with a permanent magnet (Fig. 9). Further investigations by TEM or scattering techniques remain necessary to describe the structure of membranes at a higher resolution. In DICE, W droplets are thought to be covered by just the appropriate amount of copolymer to build a membrane (after complete drying of the remaining solvent). The spherical caps observed on Fig. 9 clearly point toward higher $B$ field intensity, and thus are more likely an excess of MNPs rather than of copolymer.

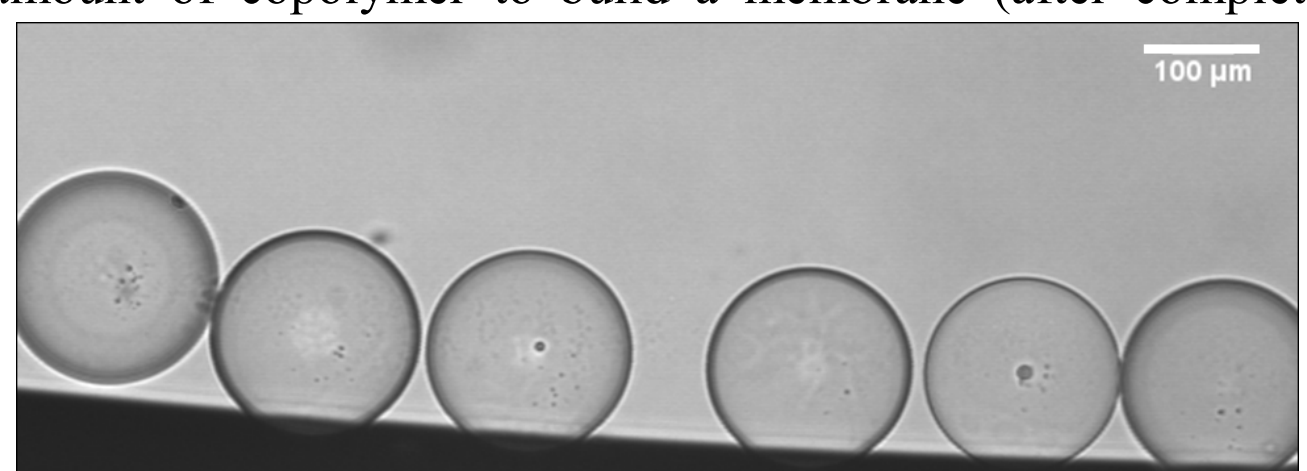

Fig. 8. Giant magnetic polymer vesicles (polymersomes) prepared with PBD$_{6.5k}$-*b*-PEO$_{3.9k}$ and an oleic acid magnetic fluid in a double-emulsion co-flows capillary device combined with the magnetic-assisted shearing-off technique. The vesicles look very regular in sizes and aspect with a diameter $D$=167±5 µm pretty close to the initial diameter $D_{\text{Int}}$ of the inner W droplets and a ratio of standard deviation to mean (coefficient of variation) CV≈3%.

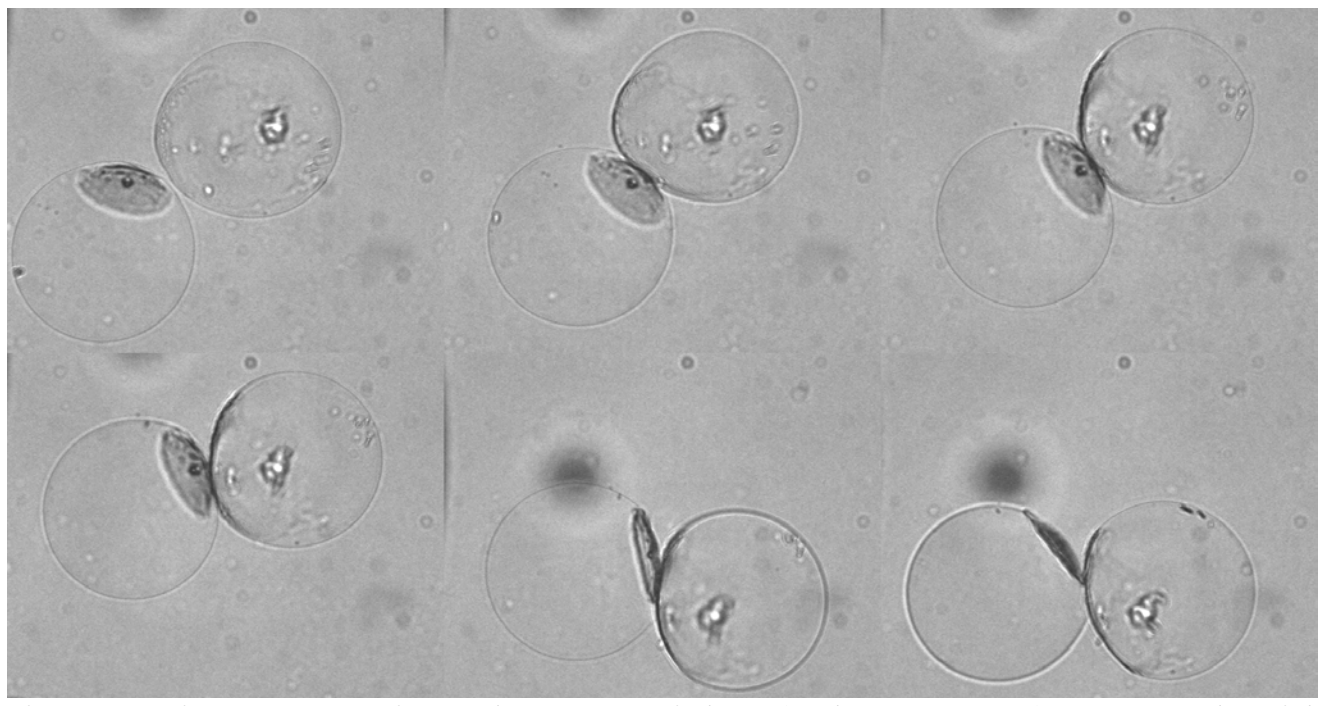

Fig. 9. Giant magnetic polymer vesicles (polymersomes) prepared with PBD$_{6.5k}$-*b*-PEO$_{3.9k}$ and an oleic acid magnetic fluid in a double-emulsion co-flows capillary device combined with magnetic-assisted shearing-off. When applying a magnetic field with a permanent magnet (thus also a gradient), the vesicle undergoes solid rotation due to attraction of the acorn-like structure.

The observation of giant magnetic polymersomes based on PBD$_{6.5k}$-*b*-PEO$_{3.9k}$ under magnetic field did not show any ellipsoidal deformation like expected. The reason might be the viscoelastic behavior of the PBD block, with a molar mass above the entanglement mass. Thus magnetic polymersomes were prepared from a 25:75 (molar) mixture of PBD$_{6.5k}$-*b*-PEO$_{3.9k}$ and PBD$_{2.5k}$-*b*-PEO$_{1.3k}$. The vesicle on Fig. 10 actually shows an ellipsoidal deformation when it is attracted by the gradient of a permanent magnet. The experiment has now to be confirmed under a homogeneous field (*i.e.* Helmholtz coil).

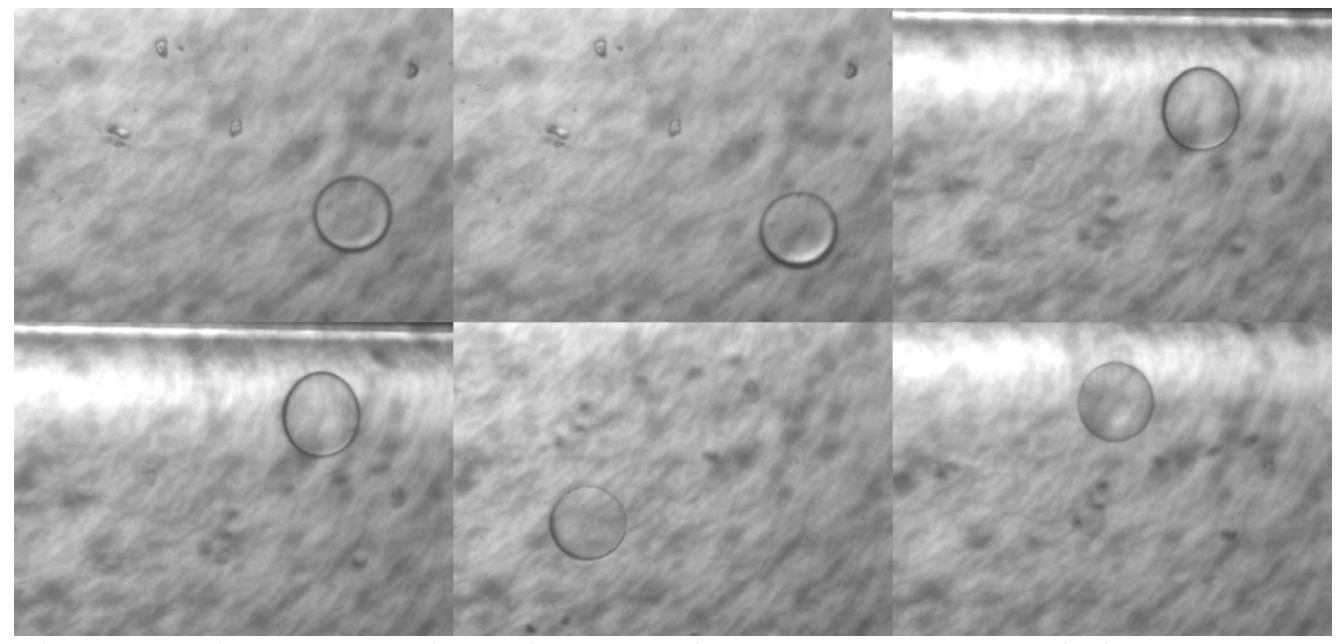

Fig. 10. Giant magnetic polymer polymersomes prepared from a 25:75 mixture of PBD$_{6.5k}$-*b*-PEO$_{3.9k}$ and PBD$_{2.5k}$-*b*-PEO$_{1.3k}$ observed under magnetic field in a capillary of rectangular section 200µm×2mm (Vitrocom). The vesicle elongated into an ellipsoid when it came closer to the magnet.

### *B. Temperature assessment at 33 nL scale*

We show here preliminary results on the possibility to assess the temperature inside a train of magnetic droplets, as sketched on Fig. 11. A theoretical prediction is actually a delicate problem involving many parameters: the specific heating power of the MNPs (depending itself on the frequency and intensity of the magnetic field); the concentration of MNPs within the droplets, expressed either as weight concentration $C_{\gamma Fe2O3}$ or as volume fraction $\Phi_{\gamma Fe2O3}$; the radius $r_{drop}$ of the droplets; the mass densities $\rho_2$ and $\rho_3$ for the liquids and $\rho_{\gamma Fe2O3}$ for iron oxide; the thermal conductivity $k_{oil}$ and the specific heat capacity $C_p$ of the oil, the *ID*, *OD* and $C_p$ of the glass capillary… We derived analytical equations valid only in the limit cases of either: (1) isolated droplet in an infinite medium with finite thermal conductivity $k_{oil}$; (2) adiabatic case where all the heat produced during the residence time $\tau_{res}$ of the droplets inside the coil is transferred into a temperature increase of all the components of the system. The exact comparison between experimental data and theory is beyond the scope of this paper. Here the idea is simply to check if the magnetic heating experiment can be performed at nL scale, and if the maximal temperature increase $\Delta T_{max}$ measured at the outlet of the channel varies with the relevant parameters (iron oxide concentration, flow-rates…) in a way that can orient us towards one of the two limit cases described by the equations:

$$\Delta T_{max} = SHP \cdot C^{drop}_{\gamma Fe_2O_3} \cdot \frac{4\pi\, r^2_{drop}}{k_{oil}} \qquad (1)$$

in the heat diffusion-limited case, or:

$$\Delta T_{max} = \frac{\rho_{\gamma Fe_2O_3} \cdot Q_2 \cdot \Phi^{drop}_{\gamma Fe_2O_3}}{\rho_{\gamma Fe_2O_3} \cdot Q_2 \cdot \Phi^{drop}_{\gamma Fe_2O_3} + Q_2 \cdot \rho_2 + (Q_2 + Q_3) \cdot \rho_3} \times \frac{SHP \cdot \tau_{res}}{C^{oil}_p + C^{glass}_p \cdot \dfrac{\rho_{glass}}{\rho_3} \dfrac{OD^2 - ID^2}{ID^2}} \qquad (2)$$

in the adiabatic or convection-limited case, *i.e.* if the limiting factor is the time $\tau_{res}$ spent by the magnetic droplets in the coil.

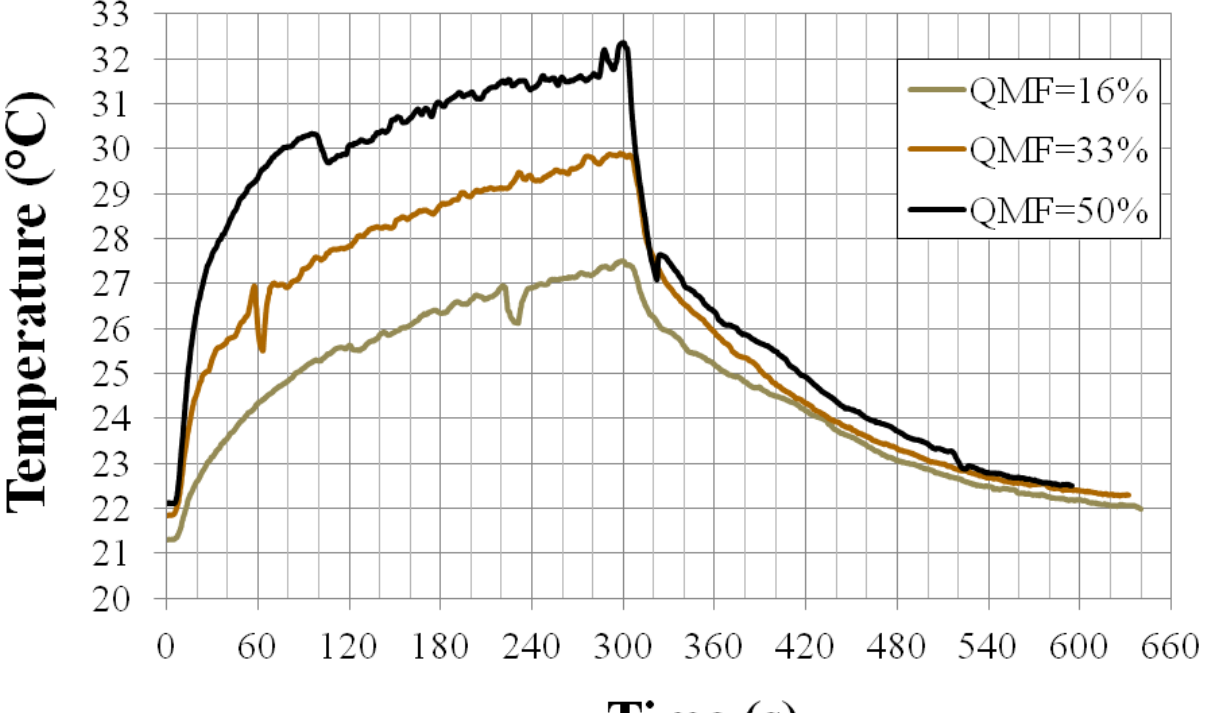

Fig. 11. Temperature measurement at the outlet of the capillary (as seen on Fig. 5) for three sets of $Q_2/Q_3$ (µL·min$^{-1}$): 24/125, 49/100, and 74/75. The total flow-rate $Q_2+Q_3$=149 µL·min$^{-1}$ and the capillary length $L$=0.65 m being kept constant, $\tau_{res}$=1 min is identical for the three experiments. Only the ratio $Q_{MF}=Q_2/(Q_2+Q_3)$ varies, giving three values of the volume fraction occupied by the train of magnetic droplets in the whole capillary, the concentration of iron oxide in the drops being also constant: $C_{\gamma Fe2O3}$=80 g·L$^{-1}$ or $\Phi_{\gamma Fe2O3}$=1.6%. The plateau values of curves $\Delta T_{max}$ are respectively: 6.2°C, 8°C and 10.3°C.

An example of experiment that can be done is given on Fig. 11. After switching off the field, the three curves superimpose, representing the kinetics of cooling down by circulation of cold liquid in the coil. But the plateau values $\Delta T_{max}$ are very different in the three cases, as ascribed to variation both in the sizes of the droplets ($r_{drop}$) and in their volume fraction $Q_{MF}$ relatively to the oil. In order to distinguish between the two regimes, we calculated $\Delta T_{max}$ theoretically using the two formulas. In this range of flow-rates $Q_2$ and $Q_3$, $r_{drop}$~200 µm but unfortunately we cannot measure it more precisely with the camera: the use of a microscope to image the droplets at a higher magnification is indeed prohibited by eddy currents that would heat the objective (containing metallic parts) in the RF magnetic field of the coil. With this approximate value of $r_{drop}$, Eq. (1) gives an estimate $\Delta T_{max}$≈6.3°C very close to the experimental values (Fig. 11 caption). To compute Eq. (2) instead, one needs first to estimate the equivalent heat capacity $C^{eq}_p = C^{oil}_p + C^{glass}_p \cdot \rho_{glass} \cdot (OD^2 - ID^2) / \rho_3 / ID^2 = 1460\ JK^{-1}kg^{-1}$ of the filled capillary with the $C_p$ and mass density $\rho_3$ of $C_6F_{14}$ given in Materials and Methods for the fluorinated oil, and the tabulated values $\rho_2$=1330 kg·m$^{-3}$ for the suspending liquid of the MF ($CH_2Cl_2$), $C_p$=840 J·kg$^{-1}$·K$^{-1}$ and $\rho_{glass}$=2100 kg·m$^{-3}$ for glass. We obtained for the three values of $Q_{MF}$=16%, 33% and 50% respectively $\Delta T_{max}$≈2.5°C, 4.5°C and 6.1°C. These values calculated by Eq. (2) are underestimating the experimental data: $\Delta T_{max}$≈6.2°C, 8°C and 10.3°C (Fig. 11). To ensure that Eq. (1) definitively describes the physical phenomena better than Eq. (2), we plotted on Fig. 12 the radius of the droplets calculated from the experimental $\Delta T_{max}$ with the heat diffusion model *vs.* the ratio of flow-rates of oil to MF ($Q_3/Q_2$). The data points are well fitted by a power law of exponent -0.154 that nicely compares to the value -0.145 reported by A. Perro *et al.* with an analogous flow focusing device [46]. Even though complications may arise compared to the simple isolated sphere model described by Eq. (1) – such as the enhanced heat transfer ascribed to the black-flows induced by the circulation of the droplets in the fluorinated oil [50] – we can nevertheless conclude that the RF induction heating of a train of magnetic droplets of radii ~200 µm in a channel correctly describes the phenomenon of MFH over a volume ~33 nL. Our study also shows the crucial need of lowering the thermal conductibility around a tumor when developing cancer treatments by MFH.

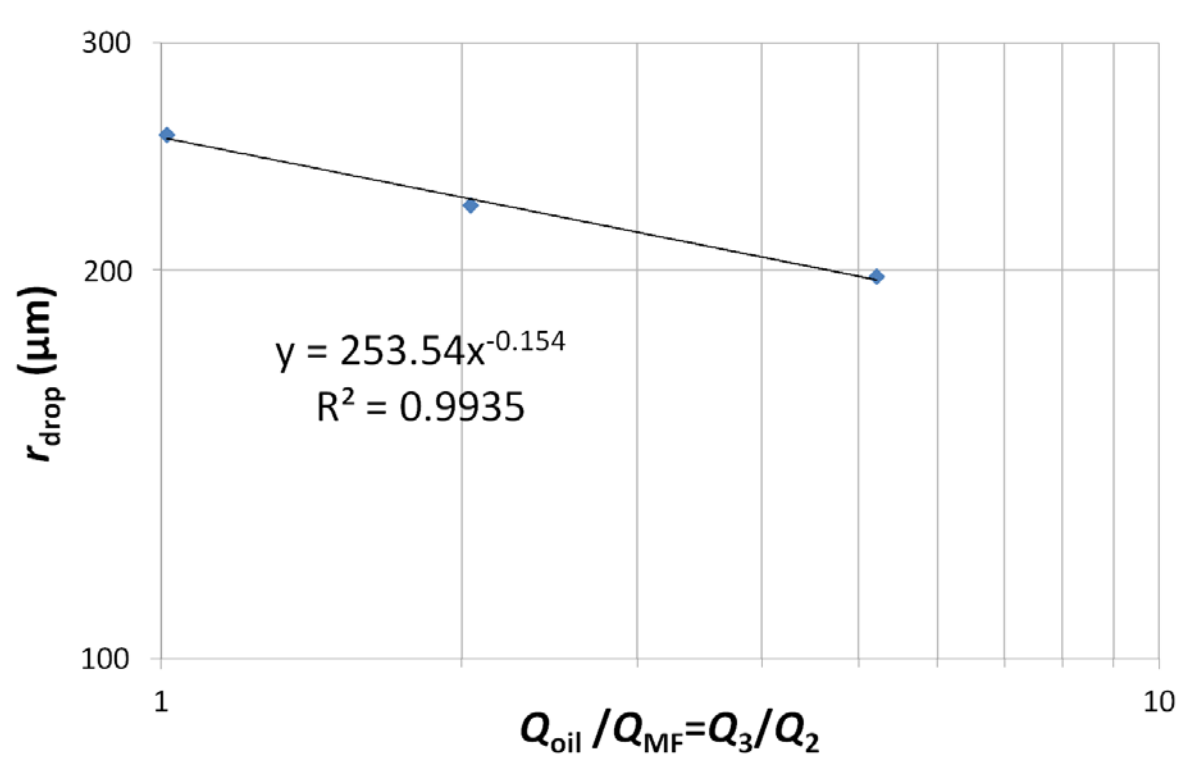

Fig. 12. Radius of the MF droplets calculated by the heat-diffusion Eq. (1) for the three ratios $Q_3/Q_2$: 125/24, 100/49, and 75/74. The data points are well fitted by a power law of exponent -0.154.

## GLOSSARY

*ID*, inner diameter; *OD*, outer diameter; SC, small capillary; MD, medium capillary; LC, large capillary; PBD, poly(butadiene); PEO, poly(ethylene oxide); PLA, poly(lactic acid); PVA, poly(vinyl alcohol); PGA, poly(L-glutamic acid); PS, poly(styrene); PTMC, poly(trimethylenecarbonate); PDMS, poly(dimethylsiloxane); DOX, doxorubicin; BSA, bovine serum albumin; OTS, octadecyltrichlorosilane; DICE, droplet interface crossing encapsulation; REV, reverse-phase evaporation; THF, tetrahydrofuran; ESF, European science foundation; P2M, precision polymer materials; DLS, dynamic light scattering; PDI, polydispersity index; TEM, transmission electron microscopy; SANS, small angle neutron scattering; MRI, magnetic resonance imaging; VSM, vibrating sample magnetometry; MF, magnetic fluid; MFH, magnetic fluid hyperthermia; RF, radiofrequency; SHP, specific heating power; MNPs, magnetic nanoparticles; CV, coefficient of variation; CNRS, National Center for Scientific Research; MI, interdisciplinarity mission; G3N, graphene, new paradigms, nanomedicine, nanometrology.

## ACKNOWLEDGMENT

This work was supported in part by the European Commission under the seventh framework by the NanoTher project (Integration of novel NANOparticle based technology for THERapeutics and diagnosis of different types of cancer CP-IP 213631-2). The authors also acknowledge ESF P2M and CNRS-MI G3N (“NanoBlast”) programs for support.

## SUPPORTING INFORMATION

Movies of the double-emulsion droplets and of the magnetic polymer vesicles are available on zenodo.org/record/4409657.

## REFERENCES

[1] W. Helfrich, "Lipid bilayer spheres: Deformation and birefringence in magnetic fields", *Physics Letters A,* vol. 43, pp. 409-410, 1973.

[2] J.-C. Bacri, V. Cabuil, A. Cebers, C. Ménager, and R. Perzynski, "Flattening of ferro-vesicle undulations under a magnetic field", *Europhysics Letters,* vol. 33, pp. 235-240, 1996.

[3] E. Alphandéry, S. Faure, O. Seksek, F. Guyot, and I. Chebbi, "Chains of Magnetosomes Extracted from AMB-1 Magnetotactic Bacteria for Application in Alternative Magnetic Field Cancer Therapy", *ACS Nano,* vol. 5, pp. 6279-6296, 2011.

[4] M. De Cuyper, P. Müller, H. Lueken, and M. Hodenius, "Synthesis of magnetic $Fe_3O_4$ particles covered with a modifiable phospholipid coat", *Journal of Physics: Condensed Matter,* vol. 15, p. S1425, 2003.

[5] O. Sandre, C. Ménager, J. Prost, V. Cabuil, J. C. Bacri, and A. Cebers, "Shape transitions of giant liposomes induced by an anisotropic spontaneous curvature", *Physical Review E,* vol. 62, pp. 3865-3870, 2000.

[6] C. Wilhelm, A. Cebers, J. C. Bacri, and F. Gazeau, "Deformation of intracellular endosomes under a magnetic field", *European Biophysics Journal,* vol. 32, pp. 655-660, 2003.

[7] J.-F. Le Meins, O. Sandre, and S. Lecommandoux, "Recent trends in the tuning of polymersomes' membrane properties (Colloquium paper)", *European Physical Journal E,* vol. 34, pp. 14(1)-14(17), 2011.

[8] S. Lecommandoux, O. Sandre, F. Checot, and R. Perzynski, "Smart hybrid magnetic self-assembled micelles and hollow capsules", *Progress in Solid State Chemistry,* vol. 34, pp. 171-179, 2006.

[9] S. Lecommandoux, O. Sandre, F. Checot, J. Rodriguez-Hernandez, and R. Perzynski, "Self-assemblies of magnetic nanoparticles and di-block copolymers: Magnetic micelles and vesicles", *Journal of Magnetism and Magnetic Materials,* vol. 300, pp. 71-74, 2006.

[10] S. Lecommandoux, O. Sandre, F. Checot, J. Rodriguez-Hernandez, and R. Perzynski, "Magnetic nanocomposite micelles and vesicles", *Advanced Materials,* vol. 17, pp. 712-718, 2005.

[11] C. Sanson, O. Diou, J. Thévenot, E. Ibarboure, A. Soum, A. Brûlet, S. Miraux, E. Thiaudière, S. Tan, A. Brisson, V. Dupuis, O. Sandre, and S. Lecommandoux, "Doxorubicin Loaded Magnetic Polymersomes: Theranostic Nanocarriers for MR Imaging and Magneto-Chemotherapy", *ACS Nano,* vol. 5, pp. 1122-1140, 2011.

[12] Y. L. Raikher and O. V. Stolbov, "Magnetodeformational effect in ferrogel objects", *Journal of Magnetism and Magnetic Materials,* vol. 289, pp. 62-65, 2005.

[13] O. V. Stolbov and Y. L. Raikher, "Deformation of a ferrovesicle in a uniform magnetic field", *Journal of Magnetism and Magnetic Materials,* vol. 300, pp. e199-e202, 2006.

[14] O. V. Stolbov, Y. L. Raikher, and M. Balasoiu, "Modelling of magnetodipolar striction in soft magnetic elastomers", *Soft Matter,* vol. 7, pp. 8484-8487, 2011.

[15] R. J. Hickey, A. S. Haynes, J. M. Kikkawa, and S.-J. Park, "Controlling the Self-Assembly Structure of Magnetic Nanoparticles and Amphiphilic Block-Copolymers: From Micelles to Vesicles", *Journal of the American Chemical Society,* vol. 133, pp. 1517-1525, 2011.

[16] G. Beaune, B. Dubertret, O. Clément, C. Vayssettes, V. Cabuil, and C. Ménager, "Giant Vesicles Containing Magnetic Nanoparticles and Quantum Dots: Feasibility and Tracking by Fiber Confocal Fluorescence Microscopy", *Angewandte Chemie International Edition,* vol. 46, pp. 5421-5424, 2007.

[17] M. I. Angelova, S. Soléau, P. Méléard, J. F. Faucon, and P. Bothorel, "Preparation of giant vesicles by external AC electric fields. Kinetics and applications", *Progress in Colloid and Polymer Science,* vol. 89, pp. 127-131, 1992.

[18] S. Pautot, B. J. Frisken, and D. A. Weitz, "Production of Unilamellar Vesicles Using an Inverted Emulsion", *Langmuir,* vol. 19, pp. 2870-2879, 2003.

[19] R. C. Hayward, A. S. Utada, N. Dan, and D. A. Weitz, "Dewetting Instability during the Formation of Polymersomes from Block-Copolymer-Stabilized Double Emulsions", *Langmuir,* vol. 22, pp. 4457-4461, 2006.

[20] H. C. Shum, J.-W. Kim, and D. A. Weitz, "Microfluidic Fabrication of Monodisperse Biocompatible and Biodegradable Polymersomes with Controlled Permeability", *Journal of the American Chemical Society,* vol. 130, pp. 9543-9549, 2008.

[21] M. Abkarian, E. Loiseau, and G. Massiera, "Continuous droplet interface crossing encapsulation (cDICE) for high throughput monodisperse vesicle design", *Soft Matter,* vol. 7, pp. 4610-4614, 2010.

[22] E. Lorenceau, A. S. Utada, D. R. Link, G. Cristobal, M. Joanicot, and D. A. Weitz, "Generation of polymerosomes from double-emulsions", *Langmuir,* vol. 21, pp. 9183-9186, 2005.

[23] J. Thiele, A. R. Abate, H. C. Shum, S. Bachtler, S. Forster, and D. A. Weitz, "Fabrication of Polymersomes using Double-Emulsion Templates in Glass-Coated Stamped Microfluidic Devices", *Small,* vol. 6, pp. 1723-1727, 2010.

[24] H. C. Shum, E. Santanach-Carreras, J.-W. Kim, A. Ehrlicher, J. Bibette, and D. A. Weitz, "Dewetting-Induced Membrane Formation by Adhesion of Amphiphile-Laden Interfaces", *Journal of the American Chemical Society,* vol. 133, pp. 4420-4426, 2011.

[25] H. C. Shum, Y. J. Zhao, S. H. Kim, and D. A. Weitz, "Multicompartment polymersomes from double emulsions", *Angewandte Chemie - International Edition,* vol. 50, pp. 1648-1651, 2011.

[26] A. Seth, G. Béalle, E. Santanach-Carreras, A. Abou-Hassan, and C. Ménager, "Design of Vesicles Using Capillary Microfluidic Devices: From Magnetic to Multifunctional Vesicles", *Advanced Materials,* vol. 24, pp. 3544-3548, 2012.

[27] A. Halbreich, J. Roger, J.-N. Pons, D. Geldwert, M.-F. Da Silva, M. Roudier, and J.-C. Bacri, "Biomedical applications of magnemite ferrofluid", *Biochimie,* vol. 80, pp. 379-390, 1998.

[28] K. Maier-Hauff, F. Ulrich, D. Nestler, H. Niehoff, P. Wust, B. Thiesen, H. Orawa, V. Budach, and A. Jordan, "Efficacy and safety of intratumoral thermotherapy using magnetic iron-oxide nanoparticles combined with external beam radiotherapy on patients with recurrent glioblastoma multiforme", *Journal of Neuro-Oncology,* vol. 103, pp. 317-324, 2011.

[29] M. Babincová, P. Cicmanec, V. Altanerová, C. Altaner, and P. Babinec, "AC-magnetic field controlled drug release from magnetoliposomes: design of a method for site-specific chemotherapy", *Bioelectrochemistry,* vol. 55, pp. 17-19, 2002.

[30] Y. Chen, A. Bose, and G. D. Bothun, "Controlled Release from Bilayer-Decorated Magnetoliposomes via Electromagnetic Heating", *ACS Nano,* vol. 4, pp. 3215–322, 2010.

[31] E. Amstad, J. Kohlbrecher, E. Müller, T. Schweizer, M. Textor, and E. Reimhult, "Triggered Release from Liposomes through Magnetic Actuation of Iron Oxide Nanoparticle Containing Membranes", *Nano Letters,* vol. 11, pp. 1664-1670, 2011.

[32] S. Nappini, M. Bonini, F. Ridi, and P. Baglioni, "Structure and permeability of magnetoliposomes loaded with hydrophobic magnetic nanoparticles in the presence of a low frequency magnetic field", *Soft Matter,* vol. 7, pp. 4801-4811, 2011.

[33] M. Dvolaitzky, P.-G. de Gennes, M.-A. Guedeau-Boudeville, and L. Jullien, "A molecular drill?", *Comptes Rendus de Académie des Sciences. Série 2,* vol. 316, pp. 1687-1690, 1993.

[34] S. Brulé, M. Levy, C. Wilhelm, D. Letourneur, F. Gazeau, C. Ménager, and C. Le Visage, "Doxorubicin Release Triggered by Alginate Embedded Magnetic Nanoheaters: A Combined Therapy", *Advanced Materials,* vol. 23, pp. 787-790, 2011.

[35] A. M. Hawkins, C. E. Bottom, Z. Liang, D. A. Puleo, and J. Z. Hilt, "Magnetic Nanocomposite Sol–Gel Systems for Remote Controlled Drug Release", *Advanced Healthcare Materials,* vol. 1, pp. 96-100, 2011.

[36] M. Creixell, A. C. Bohorquez, M. Torres-Lugo, and C. Rinaldi, "EGFR-Targeted Magnetic Nanoparticle Heaters Kill Cancer Cells without a Perceptible Temperature Rise", *ACS Nano,* vol. 5, pp. 7124-7129, 2011.

[37] R. Hergt, W. Andra, C. G. d'Ambly, I. Hilger, W. A. Kaiser, U. Richter, and H. G. Schmidt, "Physical limits of hyperthermia using magnetite fine particles", *IEEE Transactions on Magnetics,* vol. 34, pp. 3745-3754, 1998.

[38] J.-P. Fortin, C. Wilhelm, J. Servais, C. Ménager, J.-C. Bacri, and F. Gazeau, "Size-Sorted Anionic Iron Oxide Nanomagnets as Colloidal Mediators for Magnetic Hyperthermia", *Journal of the American Chemical Society* vol. 129, pp. 2628-2635, 2007.

[39] K. D. Bakoglidis, K. Simeonidis, D. Sakellari, G. Stefanou, and M. Angelakeris, "Size-Dependent Mechanisms in AC Magnetic Hyperthermia Response of Iron-Oxide Nanoparticles", *IEEE Transactions on Magnetics,* vol. 48, pp. 1320-1323, 2012.

[40] Y. Rabin, "Is intracellular hyperthermia superior to extracellular hyperthermia in the thermal sense?", *Internal Journal of Hyperthermia,* vol. 18, pp. 194-202, 2002.

[41] E. Natividad, M. Castro, and A. Mediano, "Adiabatic magnetothermia makes possible the study of the temperature dependence of the heat dissipated by magnetic nanoparticles under alternating magnetic fields", *Applied Physics Letters,* vol. 98, pp. 243119-3, 2011.

[42] R. Massart, "Preparation of aqueous magnetic liquid in alkaline and acidic media", *IEEE Transactions on Magnetics,* vol. 17, p. 1247, 1981.

[43] R. W. Chantrell, J. Popplewell, and S. W. Charles, "Measurements of particle size distribution parameters in ferrofluids", *IEEE Transactions on Magnetics,* vol. MAG-14, pp. 975-977, 1978.

[44] A. Abou Hassan, O. Sandre, V. Cabuil, and P. Tabeling, "Synthesis of iron oxide nanoparticles in a microfluidic device: preliminary results in a coaxial flow millichannel", *Chemical Communications,* pp. 1783-1785, 2008.

[45] S. Desportes, Z. Yatabe, S. Baumlin, V. Génot, J.-P. Lefèvre, H. Ushiki, J. A. Delaire, and R. B. Pansu, "Fluorescence lifetime imaging microscopy for in situ observation of the nanocrystallization of rubrene in a microfluidic set-up", *Chemical Physics Letters,* vol. 446, pp. 212-216, 2007.

[46] A. Perro, C. Nicolet, J. Angly, S. Lecommandoux, J. F. Le Meins, and A. Colin, "Mastering a Double Emulsion in a Simple Co-Flow Microfluidic to Generate Complex Polymersomes", *Langmuir,* vol. 27, pp. 8595-9068, 2011.

[47] J.-B. Brzoska, N. Shahidzadeh, and F. Rondelez, "Evidence of a transition temperature for the optimum deposition of grafted monolayer coatings", *Nature,* vol. 360, pp. 719-721, 1992.

[48] N. Pamme and C. Wilhelm, "Continuous sorting of magnetic cells via on-chip free-flow magnetophoresis", *Lab on a Chip,* vol. 6, pp. 974-980, 2006.

[49] C. B. Kristalyn, S. Watt, S. A. Spanninga, R. A. Barnard, K. Nguyen, and Z. Chen, "Investigation of sub-monolayer, monolayer, and multilayer self-assembled semifluorinated alkylsilane films", *Journal of Colloid and Interface Science,* vol. 353, pp. 322-330, 2011.

[50] Z. Che, T. N. Wong, and N.-T. Nguyen, "Heat transfer enhancement by recirculating flow within liquid plugs in microchannels", *International Journal of Heat and Mass Transfer,* vol. 55, pp. 1947-1956, 2012.